%% file: modified_SS_VB.tex
\documentclass[prd,aps,twocolumn,showpacs,amsmath,amssymb]{revtex4-2}
\usepackage{subfigure}
\usepackage{graphicx}
\usepackage{graphicx}
\usepackage{array}
\usepackage{xcolor}

\usepackage{lineno}
\usepackage{bm}
\usepackage{multirow}
\usepackage{makecell}
\usepackage[colorlinks,linkcolor=blue,anchorcolor=blue,citecolor=blue,urlcolor=blue]{hyperref}

\newcommand{\BESIIIorcid}[1]{\href{https://orcid.org/#1}{\hspace*{0.1em}\raisebox{-0.45ex}{\includegraphics[width=1em]{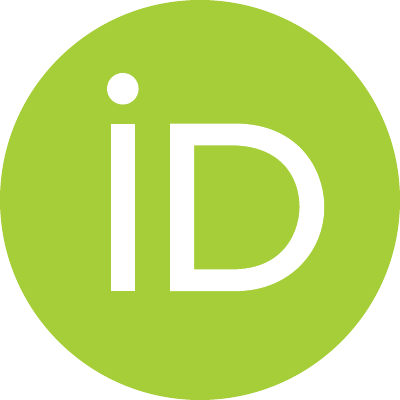}}}}

\begin{document}
\normalsize
\parskip=5pt plus 1pt minus 1pt


\title{Measurement of the branching fraction of $D^{*+}_{s}\to e^{+}e^{-}D^{+}_{s}$}

\author{
    \begin{small}
        \begin{center}
            \input{./authorlist_2025-08-14.tex}
        \end{center}
    \end{small}
}

\begin{abstract}

The branching fraction of the electromagnetic Dalitz decay $D^{*+}_{s}\to e^{+}e^{-}D^{+}_{s}$ is measured with an $e^{+}e^{-}$ collision data sample collected by the BESIII experiment at center-of-mass energies between 4.128 and 4.226~$\mathrm{GeV}$, corresponding to a total integrated luminosity of 7.33~$\mathrm{fb}^{-1}$. The measurement yields the branching fraction ${\mathcal{B}(D^{*+}_{s}\to e^{+}e^{-}D^{+}_{s})=(7.28\pm0.61_{\mathrm{stat}}\pm0.31_{\mathrm{syst}})\times10^{-3}}$. The result is consistent with the previous one, with a 2.5-fold improvement in precision. This provides an important input for constraining the parameters of theoretical models and for determining the absolute branching fractions of $D^{*+}_{s}\to \pi^{0}D^{+}_{s}$ and $D^{*+}_{s}\to \gamma D^{+}_{s}$, measured with a relative method.

\end{abstract}

\maketitle

\section{INTRODUCTION}

The measurement of electromagnetic (EM) Dalitz decays offers excellent opportunities to explore the structure of charmed mesons and to test the chiral perturbation theory in the flavor sector~\cite{EMD_Ds}. The conventional decays of a vector resonance ($V$) into a pseudoscalar meson ($P$) and a lepton pair, $V\to \gamma^{*}P\to l\bar{l}P$, provide a stringent test for theoretical models of the hadron structure and the interaction mechanism between photons and hadrons~\cite{EMD_LV}. To describe the interaction between photons and mesons, the vector-meson dominance (VMD) model is used and is parameterized with the parameter $\Lambda\approx m_{V'}$, where $m_{V'}$ is the mass of the dominant vector-meson mediator. This approximation is successful when predicting the EM Dalitz decays of vector mesons, including charmonia and bottomia~\cite{EMD_Ds}. 

Experimentally, the EM Dalitz decays of light vector mesons, such as ${\omega\to e^{+}e^{-}\pi^{0}}$~\cite{OmegaToPi0ee}, ${\phi\to e^{+}e^{-}\pi^{0}}$~\cite{PhiToPi0ee} and ${\phi\to e^{+}e^{-}\eta}$~\cite{PhiToEtaee}, have been studied widely. Some charmonium EM Dalitz decays have also been reported, such as $J/\psi\to e^{+}e^{-}\pi^{0}$~\cite{JpsiToPee}, $J/\psi\to e^{+}e^{-}\eta$~\cite{JpsiToPee} and $\psi(3686)\to e^{+}e^{-}\eta'$~\cite{Psi2SToEtapee}. Previously, the EM Dalitz decays of charmed mesons $D^{*+}_{s}\to e^{+}e^{-}D^{+}_{s}$~\cite{CELO_DsEE} and $D^{*0}\to e^{+}e^{-}D^{0}$~\cite{BESIII_D0EE} have been studied by the CLEO-c and BESIII experiments, respectively.

According to the VMD model, the differential decay width of the $D^{*+}_{s}\to e^{+}e^{-}D^{+}_{s}$ process is given by~\cite{EMD_Ds}
\begin{equation} 
	\label{eq:diff_decaywidth}
	\begin{aligned}
            d\Gamma &= \frac{1}{3}\frac{\alpha f^{2}(q^{2})}{(4\pi)^{4}q^{2}}\sqrt{1-\frac{4m_{e}^{2}}{q^{2}}}p^{3}_{D_{s}} \\ &\times\left[\left(1 +\frac{4m_{e}^{2}}{q^{2}}\right) + \left(1-\frac{4m_{e}^{2}}{q^{2}}\right)\mathrm{cos}^{2}\theta^{*}\right]dq^{2}d\Omega^{*}d\Omega,
	\end{aligned}
\end{equation}
where $\alpha\approx 1/137$ is the fine structure constant, $q$ is the four-momentum of the dilepton pair, $m_{e}$ is the electron mass, 
$\theta^{*}$ and $\Omega^{*}$ are the polar angle and the solid angle of the positive lepton in the rest frame of the virtual photon, respectively, and $\Omega$ is the solid angle of the $D_{s}^{+}$ meson in the $D_{s}^{*+}$ rest frame. The term $p_{D_{s}}$ is the momentum of the $D^{+}_{s}$ in the $D^{*+}_{s}$ rest frame, and is given by
\begin{equation} 
	\label{eq:diff_pDs}
	\begin{aligned}
            p_{D_{s}}=\frac{\sqrt{\lambda(m^{2}_{D^{*}_{s}},m^{2}_{D_{s}},q^{2})}}{2m_{D^{*}_{s}}},
	\end{aligned}
\end{equation}
where the term $\lambda(a,b,c)=a^{2}+b^{2}+c^{2}-2ab-2bc-2ac$ represents the Kallen function. The form factor $f(q^{2})$ at the simple pole approximation is given by
\begin{equation}
	\label{eq:simFF}
	f(q^{2})=\frac{f(0)}{1-q^{2}/m_{V'}^{2}},
\end{equation}
\noindent
where $f(0)$ is the form factor of the $D_{s}^{*+}\to\gamma D_{s}^{+}$ decay. 

Similarly, the $D_{s}^{*+}\to\gamma D_{s}^{+}$ decay width is given by
\begin{equation}
	\label{eq:gammaDs}
	\Gamma_{\gamma D_{s}^{+}}=\frac{f^{2}(0)}{12\pi}\left(\frac{m^{2}_{D_{s}^{*+}}-m^{2}_{D_{s}^{+}}}{2m_{D_{s}^{*+}}}\right)^{3},
\end{equation}
\noindent
where $m_{D_{s}^{+}}$ and $m_{D_{s}^{*+}}$ are the masses of $D_{s}^{+}$ and $D_{s}^{*+}$, respectively. The relative branching fraction of $D^{*+}_{s}\to e^{+}e^{-}D^{+}_{s}$ to $D^{*+}_{s}\to \gamma D^{+}_{s}$ is $6.46\times10^{-3}$, derived by integrating the differential relative decay width over the phase space~\cite{EMD_Ds}, where $m_{V'}$ is set at the $\phi$ resonance mass and all masses are quoted from the Particle Data Group (PDG)~\cite{PDG2023}.

The branching fraction of the decay $D^{*+}_{s}\to e^{+}e^{-}D^{+}_{s}$ is a key input for constraining the parameters of theoretical models, and for determining the absolute branching fractions of the $D^{*+}_{s}\to \pi^{0}D^{+}_{s}$ and $D^{*+}_{s}\to \gamma D^{+}_{s}$ decays when a relative normalization approach is adopted. Therefore, improving the precision of this measurement is highly motivated. The BESIII experiment has collected a large electron-positron collision data sample with center-of-mass (CM) energies ($\sqrt{s}$) ranging between 4.128 and 4.226~$\mathrm{GeV}$, corresponding to a total integrated luminosity of 7.33~$\mathrm{fb}^{-1}$~\cite{BESIII2021_A,BESIII2022_A,BESIIIFuture_A}. The process $e^{+}e^{-}\to D_{s}^{*}D_{s}$ is abundantly produced near the $D_{s}^{*}D_{s}$ production threshold~\cite{BESIII2021_A,BESIII2022_A,BESIIIFuture_A}, providing an excellent opportunity to measure the branching fraction of $D^{*+}_{s}\to e^{+}e^{-}D^{+}_{s}$ using a tagging technique~\cite{MarkIII:DT,Ke:2023qzc}. Throughout this manuscript, $D_{s}^{*}D_{s}$ denotes $D_{s}^{*+}D_{s}^{-}$ and $D_{s}^{*-}D_{s}^{+}$.

\section{BESIII DETECTOR AND MONTE CARLO SIMULATION}

The BESIII detector~\cite{Ablikim:2009aa} records data from the symmetric $e^{+}e^{-}$ collisions provided by the BEPCII storage ring~\cite{Ring:2016} in the CM energy ranging from 1.84~GeV to 4.95~GeV, with a peak luminosity of ${1.1\times 10^{33}~\mathrm{cm}^{-2}\mathrm{s}^{-1}}$ achieved at 3.773~GeV. BESIII has collected a large data sample in this energy region~\cite{BESIII2020a,EcmsMea,EventFilter}. The cylindrical core of the BESIII detector covers 93\% of the full solid angle and consists of a helium-based multilayer drift chamber (MDC), a plastic scintillator time-of-flight system (TOF), and a CsI(Tl) electromagnetic calorimeter (EMC). All these detectors are enclosed in a superconducting solenoidal magnet providing a 1.0~T magnetic field. The solenoid is supported by an octagonal flux-return yoke with resistive plate counter (RPC) of muon-identification (MUC) modules interleaved with steel. The main function of the MUC is to separate muons from charged pions and other hadrons based on their hit patterns in the flux return yoke. The MDC provides the charged-particle momentum with a resolution of 0.5\% at 1~$\mathrm{GeV}/c$, and the specific ionization energy loss ($\mathrm{d}E/\mathrm{d}x$) resolution of 6\% for electrons from Bhabha scattering. The EMC measures photons with an energy resolution of 2.5\% (5\%) at 1~GeV in the barrel (end-cap) region. The time resolution in the TOF barrel region is 68~ps. That in the end-cap region was 110~ps before the end-cap TOF system was upgraded in 2015 using multigap resistive plate chamber technology, providing a time resolution of 60~ps, which benefits 84\% of the data used in this analysis~\cite{etofa,etofb,etofc}.
Data samples used in this work are summarized in Table~\ref{tab:alllum}, including those with CM energies ranging between 4.128 and 4.226~GeV, corresponding to a total integrated luminosity of 7.33~$\mathrm{fb}^{-1}$.

\begin{table}[htbp]
	\centering
	\caption{Summary of integrated luminosities ($\mathcal{L}_{\mathrm{int}}$) and requirements on $M_{\mathrm{rec}}$ for different CM energies. The first and second uncertainties are statistical and systematic, respectively. The definition of $M_{\mathrm{rec}}$ is in Eq.~\ref{eq:MrecDs}. The integrated luminosities for data sample at $\sqrt{s}=4.128~\mathrm{GeV}$ and 4.157~GeV are estimated by using the online monitoring information.} 
	\label{tab:alllum}	
	\begin{tabular}[b]{ p{1.5cm} p{2.0cm}  p{2.0cm} }
		\hline
		\hline
		\multicolumn{1}{l}{\small{$\sqrt{s}$ (GeV)}} &  \multicolumn{1}{c}{\small{~~$\mathcal{L}_{\mathrm{int}}~(\mathrm{pb}^{-1})$~\cite{BESIII2021_A,BESIII2022_A,BESIIIFuture_A}~~}} & \multicolumn{1}{c}{\small{~~~~$M_{\mathrm{rec}}~(\mathrm{GeV}/c^{2})$~\cite{BESIII:ABrDs}~~~~}} \\
		\hline
		$4.128$ & \multicolumn{1}{c}{$398.4\pm3.2$} & \multicolumn{1}{c}{$[2.060,~2.150]$} \\
		
		$4.157$ & \multicolumn{1}{c}{$411.7\pm3.3$} & \multicolumn{1}{c}{$[2.054,~2.170]$} \\
		
		$4.178$ & \multicolumn{1}{c}{$3189.0\pm0.2\pm31.9$} & \multicolumn{1}{c}{$[2.050,~2.180]$} \\
		
		$4.189$ & \multicolumn{1}{c}{$570.0\pm0.1\pm2.2$} & \multicolumn{1}{c}{$[2.048,~2.190]$} \\
		
		$4.199$ & \multicolumn{1}{c}{$526.0\pm0.1\pm2.1$} & \multicolumn{1}{c}{$[2.046,~2.200]$} \\
		
		$4.209$ & \multicolumn{1}{c}{$572.1\pm0.1\pm1.8$} & \multicolumn{1}{c}{$[2.044,~2.210]$} \\
		
		$4.219$ & \multicolumn{1}{c}{$569.2\pm0.1\pm1.8$} & \multicolumn{1}{c}{$[2.042,~2.220]$} \\
		
		$4.226$ & \multicolumn{1}{c}{$1100.9\pm0.1\pm7.0$} & \multicolumn{1}{c}{$[2.040,~2.220]$} \\
		
		\hline	
		\hline
	\end{tabular}
\end{table}

The MC simulated samples produced with a {\footnotesize{GEANT}}4-based~\cite{Geant4:2002hh} package, which includes the geometric description of the BESIII detector and the detector response, are used to model the signal, determine signal detection efficiency and to estimate backgrounds. The MC samples are produced with known decay rates~\cite{PDG2023} and correct angular distributions by two event generators, {\footnotesize{EVTGEN}}~\cite{Lange:2001uf,Ping:2008zz} for both charm ($D_{s}^{*\pm}$, $D_{s}^{\pm}$, $D^{*0(\pm)}$ and $D^{0(\pm)}$) and charmonium decays and {\footnotesize{KKMC}}~\cite{KKMC:2000,KKMC:2001} for continuum processes. The samples consist of $e^{+}e^{-}\to D\bar{D}$, $D^{*}D$, $D^{*}D^{*}$, $D_{s}D_{s}$, $D^{*}_{s}D_{s}$, $D^{*}_{s}D^{*}_{s}$, $\pi DD^{*}$, $\pi DD$,  $q\bar{q}(q=u,d,s)$, $\gamma J/\psi$, $\gamma\psi(3686)$ and $\tau^{+}\tau^{-}$. Charmonium decays that are not accounted for by exclusive measurements are simulated by {\footnotesize{LUNDCHARM}}~\cite{Chen:2000tv,Ping:2014}. All MC simulations include the effects of initial-state radiation (ISR), final-state radiation (FSR) and beam energy spread in  $e^{+}e^{-}$ annihilation, where ISR is simulated with ConExc~\cite{BESIII:ConExc} for $e^{+}e^{-}\to c\bar{c}$ events within the framework of {\footnotesize{EVTGEN}}, and with {\footnotesize{KKMC}} for non-charm continuum processes. FSR for charged final state particles is incorporated using the {\footnotesize{PHOTOS}} package~\cite{photos2}.
The beam energy spread in the $e^{+}e^{-}$ annihilation is modeled with {\footnotesize{KKMC}}.  
Signal MC samples used to estimate the signal efficiency for different tag modes are generated individually, and the decay $D_{s}^{*\pm}\to e^{+}e^{-}D_{s}^{\pm}$ follows the differential decay width given in Eq.~\ref{eq:diff_decaywidth}.

\section{DETERMINATION OF THE TAG AND SIGNAL YIELDS}

The $e^{+}e^{-}\to D_{s}^{*}D_{s}$ events are reconstructed by tagging the $D_{s}^{\pm}$ signal (denoted as $D_{s}$) with the decay modes (denoted as $f$) that have large branching fractions, and the corresponding yields are
\begin{equation}
	\label{eq:STY}
	S(f^{\pm})=2N_{D_{s}^{*}D_{s}}\mathcal{B}(f)\epsilon(f^{\pm}),
\end{equation}
\noindent
where $N_{D_{s}^{*}D_{s}}$ is the total number of $e^+e^-\to D_{s}^{*+}D_{s}^-$ or $D_{s}^{*-}D_{s}^+$ events in the data sample, the factor 2 accounts the charge conjugated modes of $e^{+}e^{-}\to D_{s}^{*}D_{s}$,
$\mathcal{B}(f)$ and $\epsilon(f^{\pm})$ are the branching fraction and reconstruction efficiency of the tag mode $D_{s}^{\pm}\to f^{\pm}$, respectively. The eleven $D_{s}^{\pm}$ tag modes employed in this analysis are summarized in Table~\ref{tab:alltagmode}, where the intermediate states are reconstructed with the following decays: $K^{0}_{S}\to \pi^{+}\pi^{-}$, $\pi^{0}\to\gamma\gamma$, $\eta\to \gamma\gamma$, $\eta'\to\pi^{+}\pi^{-}\eta$ and $\gamma\rho^{0}$ ($\rho^{0}\to\pi^{+}\pi^{-}$). The reconstructed $D_{s}^{\pm}$ mesons may originate either from $D_{s}^{*}$ decays (i.e. daughter $D_{s}^{\pm}$) or be directly produced in the $e^+e^-$ collision in association with a $D_{s}^{*}$ (i.e. bachelor $D_{s}^{\pm}$). These two contributions cannot be distinguished at the present stage.

\begin{table}[htbp]
	\centering
	\caption{Summary of the $D_s^{\pm}$ tagging modes and their branching fractions from the PDG~\cite{PDG2023}, where the quoted uncertainties include the contributions from both primary and secondary branching fractions.} 
	\label{tab:alltagmode}	
	\begin{tabular}[b]{ p{3.5cm} p{3.5cm} }
		\hline
		\hline
        \multicolumn{1}{l}{\small{Mode $f^{\pm}$}} & \multicolumn{1}{c}{\small{Effective branching fraction(\%)}} \\
		\hline
        $K_{S}^{0}K^{\pm}$ & \multicolumn{1}{c}{$1.04\pm0.01$} \\

        $K^{+}K^{-}\pi^{\pm}$ & \multicolumn{1}{c}{$5.45\pm0.08$} \\

        $K_{S}^{0}K^{\pm}\pi^{0}$ & \multicolumn{1}{c}{$1.01\pm0.02$} \\

        $K^{+}K^{-}\pi^{\pm}\pi^{0}$ & \multicolumn{1}{c}{$5.46\pm0.15$} \\

        $K_{S}^{0}K^{\mp}\pi^{\pm}\pi^{\pm}$ & \multicolumn{1}{c}{$1.09\pm0.02$} \\

        $\pi^{-}\pi^{+}\pi^{\pm}$ & \multicolumn{1}{c}{$1.09\pm0.01$} \\

        $\pi^{\pm}\eta$ & \multicolumn{1}{c}{$0.66\pm0.01$} \\

        $\pi^{\pm}\pi^{0}\eta$ & \multicolumn{1}{c}{$3.54\pm0.07$} \\

        $\pi^{\pm}\eta'_{\pi^{+}\pi^{-}\eta}$ & \multicolumn{1}{c}{$0.66\pm0.02$} \\

        $\pi^{\pm}\eta'_{\gamma\rho^{0}}$ & \multicolumn{1}{c}{$1.16\pm0.03$} \\

        $K^{\pm}\pi^{+}\pi^{-}$ & \multicolumn{1}{c}{$0.62\pm0.01$} \\

		\hline	
		\hline
	\end{tabular}
\end{table}

Charged tracks in the MDC are required to satisfy $|\mathrm{cos}\theta|<0.93$, where $\theta$ is the polar angle defined with respect to the $z$-axis, which is the symmetry axis of the MDC.
For charged tracks not originated from $K^{0}_{S}$ decay, the distance of closest approach to the interaction point (IP) must be less than 10~cm along the $z$-axis ($|V_{z}|$) and less than 1~cm in the transverse plane ($V_{xy}$). Particle identification (PID) for charged tracks is performed by combining information from the $\mathrm{d}E/\mathrm{d}x$ in the MDC and the flight time in the TOF to form likelihoods $\mathcal{L}(h)~(h=K,\pi)$ for each hadron $h$ hypothesis. Charged kaons and pions are identified by comparing the obtained likelihood values and requiring $\mathcal{L}(K)>\mathcal{L}(\pi)$ and $\mathcal{L}(\pi)>\mathcal{L}(K)$, respectively.

Photon candidates are identified using showers in the EMC. The deposited energy of each shower must be greater than 25~MeV in the barrel region ($|\mathrm{cos}\theta|<0.80$) or 50~MeV in the end-cap region ($0.86<|\mathrm{cos}\theta|<0.92$). To exclude showers originating from charged tracks, the opening angle between the EMC shower and the closest charged track at the EMC must be greater than $10^{\circ}$ as measured from the IP. To suppress electronic noise and showers unrelated to the event, the difference between the EMC time and the event start time must be within $[0,700]$~ns.

The $K^{0}_{S}$ candidates are reconstructed with two oppositely charged tracks satisfying $|V_{z}|<20~\mathrm{cm}$; the two charged tracks are assigned
as $\pi^+\pi^-$ without imposing further PID criteria. They are constrained to
originate from a common vertex and are required to have an invariant mass
within $[0.487,0.511]~\mathrm{GeV}/c^{2}$. The
decay length of the $K^0_S$ candidate is required to be greater than
twice the vertex resolution away from the IP.

The $\pi^{0}$ and $\eta$ candidates are reconstructed using photon pairs with $\gamma\gamma$ invariant masses within $[0.115,0.150]$ and $[0.490,0.580]$~$\mathrm{GeV}/c^{2}$, respectively. In order to improve the momentum resolution, a kinematic fit is carried out on the $\gamma\gamma$ candidates, where the invariant mass of the photon pair is constrained to the known $\pi^{0}$ or $\eta$ mass~\cite{PDG2023}. The kinematic variables from the fits are used in the subsequent analysis. The $\rho^{0}$ candidates are reconstructed from the decay mode $\rho^{0}\to \pi^{+}\pi^{-}$, requiring the invariant mass of the $\pi^{+}\pi^{-}$ pair to be within $[0.570,0.970]~\mathrm{GeV}/c^{2}$.
The $\eta'$ candidates are reconstructed from the decay modes $\eta'\to\pi^{+}\pi^{-}\eta$ and $\eta'\to\gamma\rho^{0}$, with the invariant mass of the $\pi^{+}\pi^{-}\eta$ and $\gamma\pi^{+}\pi^{-}$ combinations being in the ranges of $[0.943,0.973]$ and $[0.946,0.970]~\mathrm{GeV}/c^{2}$, respectively.

The single tag (ST) samples are selected by reconstructing a $D_s^{\pm}$ candidate with the selected $\pi^{\pm}$, $K^{\pm}$, $\pi^0$, $K^{0}_{S}$, $\eta$ and $\eta'$ candidates and requiring 
the invariant mass within $[1.85,~2.06]~\mathrm{GeV}/c^{2}$. 
To suppress the backgrounds from wrong combinations of $\pi^{\pm}$ and $\pi^{0}$ from the decays of $D_{s}^{*\pm}$, $D^{*0}$ and $D^{*\pm}$ to $D_{s}^{\pm}$, $D^{0}$ and $D^{\pm}$, respectively, the momenta of $\pi^{\pm}$ and $\pi^{0}$ in the ST $D_{s}^{\pm}$ candidates are required to be larger than $100~\mathrm{MeV}/c$.
For the $D_{s}^{\pm}\to K^{\pm}\pi^{+}\pi^{-}$ and $D_{s}^{\pm}\to\pi^{+}\pi^{-}\pi^{\pm}$ modes, candidates are rejected if any $\pi^{+}\pi^{-}$ combination (two combinations in the latter mode) has an invariant mass within $[0.468,~0.528]~\mathrm{GeV}/c^{2}$, to suppress the backgrounds associated to the $K_{S}^{0}$ decays. 

The recoil mass $M_{\mathrm{rec}}$ against tagged $D_{s}$ is defined as
\begin{equation}
    \label{eq:MrecDs}
    M_{\mathrm{rec}}=\sqrt{\left (\sqrt{s}-\sqrt{|\vec{p}_{\mathrm{tag}}|^{2}+m_{D_{s}^{+}}^{2}}\right )^{2}-|\vec{p}_{\mathrm{tag}}|^{2}},
\end{equation}
where $\vec{p}_{\mathrm{tag}}$ is the three-momentum of the $D_{s}^{\pm}$ candidate in the $e^{+}e^{-}$ CM frame, and $m_{D_{s}^{+}}$ is the $D_{s}^{\pm}$ nominal mass quoted from the PDG~\cite{PDG2023}. If multiple combinations are present for a specific mode, only the candidate with the closest $M_{\mathrm{rec}}$ to the $D_{s}^{*+}$ nominal mass~\cite{PDG2023} is retained for the subsequent analysis. 
The requirements of $M_{\mathrm{rec}}$ to suppress combinatorial and $e^{+}e^{-}\to D_{s}^{+}D_{s}^{-}$ backgrounds are summarized in Table~\ref{tab:alllum} for each CM energy point.

\begin{figure*}[htpb]
	\begin{center}
		\subfigure{\includegraphics[width=0.96\textwidth]{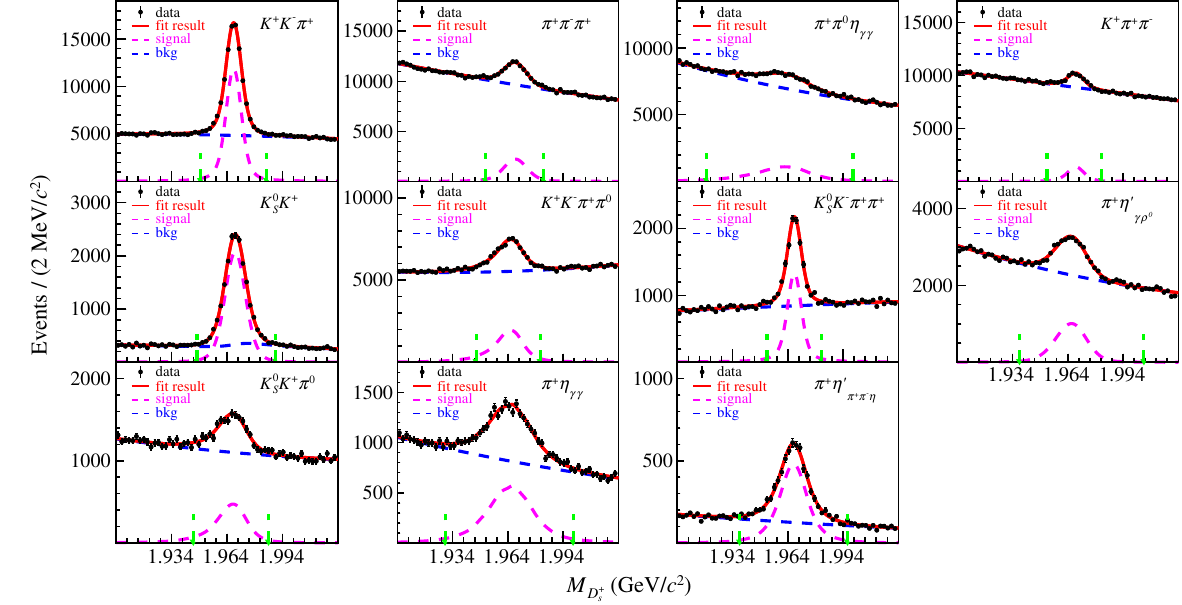}}
	\end{center}
     \vskip -0.3cm
	\setlength{\abovecaptionskip}{-0.0cm}
	\caption{
        The $M_{D_{s}^{+}}$ distributions of different tag modes for the data sample at $\sqrt{s}=4.178~\mathrm{GeV}$. The black dots with error bars are data. The red solid curves represent the fit results and the blue (pink) dashed curves describe the background (signal) shapes. The ST mass windows are between the green dashed arrows.  
	} \label{fig:ST_Dsp_4180}
\end{figure*}

\begin{figure*}[htpb]
	\begin{center}
		\subfigure{\includegraphics[width=0.96\textwidth]{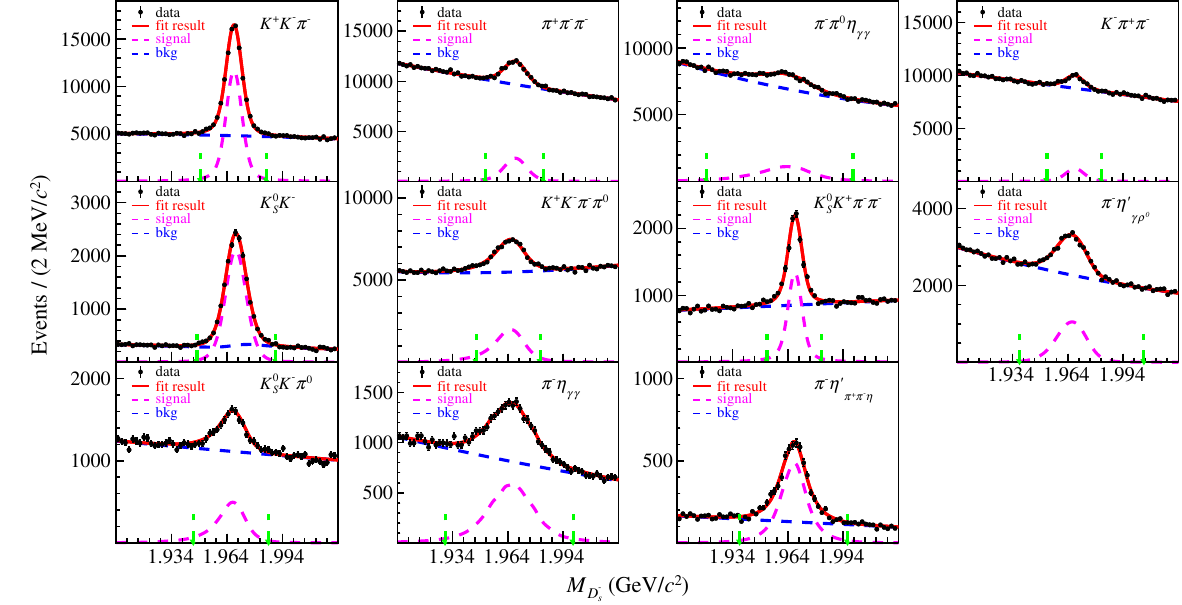}}
	\end{center}
     \vskip -0.3cm
	\setlength{\abovecaptionskip}{-0.0cm}
	\caption{
        The $M_{D_{s}^{-}}$ distributions of different tag modes for the data sample at $\sqrt{s}=4.178~\mathrm{GeV}$. The black dots with error bars are data. The red solid curves represent the fit results and the blue (pink) dashed curves describe the background (signal) shapes. The ST mass windows are between the green dashed arrows.  
	} \label{fig:ST_Dsm_4180}
\end{figure*}

The ST yields $S(f^\pm)$ are extracted by fitting the distribution of the ST $D_{s}^{\pm}$ mass, defined as $M_{D_{s}^{\pm}}=\sqrt{E_{\mathrm{tag}}^{2}-|\vec{p}_{\mathrm{tag}}|^{2}}$. The distributions are shown in Figs.~\ref{fig:ST_Dsp_4180} and~\ref{fig:ST_Dsm_4180} at $\sqrt{s}=4.178~\mathrm{GeV}$ as an example, where $E_{\mathrm{tag}}$ is the energy of the ST $D_{s}^{\pm}$ candidate obtained by summing the energies of the daughter particles in the $e^{+}e^{-}$ CM frame. 
Detailed studies based on the inclusive MC sample indicate that there is no peaking background in the $M_{D_{s}^{\pm}}$ distribution for most tag modes. Therefore, unbinned maximum likelihood fits to the $M_{D_{s}^{\pm}}$ distributions are performed for candidates with invariant masses between 1.90~GeV/$c^2$ and 2.03~GeV/$c^2$, separately for each tag modes and for data samples collected at different CM energies. 
In the fit, the signal is described by the MC simulated shape convolved with a Gaussian function, which accounts for the resolution difference between data and MC simulation, and the combinatorial background is modeled with a second order Chebychev polynomial function, which has been verified by analyzing the inclusive MC sample. In the fit of the $D_{s}^{\pm}\to K_{S}^{0}K^{\pm}$ channel, the shape of the peaking background from $D^{\pm}\to K_{S}^{0}\pi^{\pm}$ is modeled by the shape from the inclusive MC sample, and the corresponding size is left floating in the fit. 
To improve the signal-to-background ratio, reduced $D_{s}^{\pm}$ mass windows are applied for each individual decay mode.

Table~\ref{tab:STDspm} summarizes the reduced $D_{s}^{\pm}$ mass windows, as well as the obtained ST yields ($S(f^{\pm})$) and ST efficiencies ($\epsilon (f^{\pm})$), for each ST mode at $\sqrt{s}=4.178~\mathrm{GeV}$ as an example, where the detection efficiencies are obtained with the inclusive MC sample. 
The values obtained for the other CM energies are presented in the Supplemental Material~\cite{Supp}.
The sum of the ST yields across all eight data samples is $(378.8\pm2.3)\times 10^{3}$ for ST $D_{s}^{+}$ and $(383.7\pm2.3)\times 10^{3}$ for ST $D_{s}^{-}$, where the uncertainties are statistical only.

\begin{table*}[htbp]
	\centering
	\caption{Summary of the ST mass windows (in GeV/$c^2$), the corresponding tag yields ($S(f^{\pm})$) and tag efficiencies ($\epsilon(f^{\pm})$) for the different $D_{s}^{\pm}$ tag modes of the data sample with $\sqrt{s}=4.178~\mathrm{GeV}$. The uncertainties are statistical only.}
	\label{tab:STDspm}
	
	\begin{tabular}[b]{ l  c  p{1.5cm}<{\raggedleft} @{ $\pm$ } p{1.0cm}<{\raggedright} p{1.5cm}<{\raggedleft} @{ $\pm$ } p{1.0cm}<{\raggedright} p{1.5cm}<{\raggedleft} @{ $\pm$ } p{1.0cm}<{\raggedright} p{1.5cm}<{\raggedleft} @{ $\pm$ } p{1.0cm}<{\raggedright} }
		\hline
		\hline
        \makebox[0.070\textwidth][l]{\small{Mode}} & \makebox[0.150\textwidth][c]{\small{$\mathrm{Mass~window}$}} & \multicolumn{2}{c}{$S(f^+)$} & \multicolumn{2}{c}{$S(f^-)$} & \multicolumn{2}{c}{$\epsilon(f^+)~(\%)$} & \multicolumn{2}{c}{$\epsilon(f^-)~(\%)$} \\
		\hline
		$K_{S}^{0}K^{\pm}$                    & $(1.948,~1.991)$ & 15221 & 186 & 15637 & 192 & 47.79 & 0.22 & 47.26 & 0.22 \\

        $K^{+}K^{-}\pi^{\pm}$                 & $(1.950,~1.986)$ & 68315 & 436 & 69052 & 442 & 40.71 & 0.10 & 41.04 & 0.10 \\

        $K_{S}^{0}K^{\pm}\pi^{0}$             & $(1.946,~1.987)$ & 4881  & 278 & 4790  & 264 & 14.71 & 0.29 & 14.51 & 0.29 \\

        $K^{+}K^{-}\pi^{\pm}\pi^{0}$          & $(1.947,~1.982)$ & 18558 & 557 & 20011 & 596 & 11.04 & 0.12 & 10.79 & 0.12 \\

        $K_{S}^{0}K^{\mp}\pi^{\pm}\pi^{\pm}$      & $(1.953,~1.983)$ & 7110  & 160 & 7158  & 163 & 19.91 & 0.18 & 20.48 & 0.18 \\

        $\pi^{-}\pi^{+}\pi^{\pm}$             & $(1.952,~1.984)$ & 18142 & 678 & 18569 & 629 & 54.34 & 0.56 & 55.26 & 0.56 \\

        $\pi^{\pm}\eta$                       & $(1.930,~2.000)$ & 8922  & 313 & 9263  & 318 & 45.64 & 0.58 & 45.17 & 0.58 \\

        $\pi^{\pm}\pi^{0}\eta$ & $(1.920,~2.000)$ & 20919 & 1093 & 21213 & 1039 & 18.35 & 0.38 & 18.90 & 0.40 \\

        $\pi^{\pm}\eta'_{\pi^{+}\pi^{-}\eta}$ & $(1.938,~1.997)$ & 4939  & 127 & 4753  & 119 & 24.70 & 0.24 & 24.88 & 0.24 \\

        $\pi^{\pm}\eta'_{\gamma\rho^{0}}$     & $(1.938,~2.006)$ & 10998 & 427 & 11141 & 413 & 29.32 & 0.38 & 29.65 & 0.38 \\

        $K^{\pm}\pi^{+}\pi^{-}$               & $(1.953,~1.983)$ & 8332  & 467 & 8124  & 537 & 46.65 & 0.87 & 46.09 & 0.88 \\
        
		\hline	
		\hline
	\end{tabular}
\end{table*}

Based on the selected $e^+e^-\to D_{s}^{*}D_{s}$ samples, the events of interest $e^+ e^- \to D_{s}^{*}D_{s}\to e^+ e^- D_{s}^+D_{s}^-$ are selected by additionally reconstructing an $e^+e^-$ pair (hereafter denoted as $g$). Using the information from the reconstructed $D_{s}^{\pm}$ meson and the $e^+e^-$ pair, the tagged $D_{s}^{\pm}$ candidate can be identified as either the daughter or the bachelor meson.
The corresponding yields are 
\begin{equation}
\label{eq:DTY}
N(g, f^{\pm,\mathrm{dau/bac}})=N_{D_{s}^{*}D_{s}}\mathcal{B}(g)\mathcal{B}(f)\epsilon(g, f^{\pm,\mathrm{dau/bac}}),
\end{equation}
\noindent
where $\mathcal{B}(g)$ is the branching fraction of $D_{s}^{*+}\to e^{+}e^{-}D_{s}^{+}$, $\epsilon(g, f^{\pm,\mathrm{dau/bac}})$ is the corresponding detection efficiency, the superscripts ``dau" and ``bac" indicate whether the ST $D_{s}^{\pm}$ corresponds to the daughter or the bachelor particle, respectively.

The $e^{+}e^{-}$ pair is selected by reconstructing an additional electron and positron pair within the remaining charged tracks. Only the charged tracks with momentum less than $200~\mathrm{MeV}/c$ are used, due to the small phase space of the $D^{*+}_{s}\to e^{+}e^{-}D^{+}_{s}$ decay. To suppress the backgrounds due to the mis-identification of $\pi^{\pm}$/$K^{\pm}$ as $e^{\pm}$, the PID with the $\mathrm{d}E/\mathrm{d}x$ information from the MDC is performed to form likelihoods $\mathcal{L}(h)$ for the hypothesis of $K$, $\pi$ and $e$, individually, and the electron is identified by requiring $\mathcal{L}(e^{\pm})>0$, $\mathcal{L}(e^{\pm})>\mathcal{L}(K^{\pm})$ and $\mathcal{L}(e^{\pm})>\mathcal{L}(\pi^{\pm})$. To eliminate the background from $D_{s}^{*\pm}\to\gamma D_{s}^{\pm}$, where the $\gamma$ converts into an $e^+ e^-$ pair by interacting with the beam pipe, placed at a radius of $3~\mathrm{cm}$, or the MDC inner wall, with radius of $6~\mathrm{cm}$, the $\gamma$ conversion program~\cite{BESIII_D0EE} is carried out, where the variable $R_{xy}$, which represents the distance between the interaction point and the photon-conversion vertex in the $x-y$ plane, is calculated and the selection $R_{xy}<2.0~\mathrm{cm}$ is applied to suppress the corresponding background.

After applying all above selection criteria, the recoil mass of the tagged $D_{s}^{\pm}$ and the $e^+e^-$ pair is calculated, defined as
\begin{equation}
    \label{eq:MrecDsee}
    M^{\mathrm{rec}}_{e^{+}e^{-}D_{s}^{\pm}}=\sqrt{(\sqrt{s}-E_{e^{+}e^{-}D_{s}^{\pm}})^{2}-|\vec{p}_{e^{+}e^{-}D_{s}^{\pm}}|^{2}},
\end{equation}
\noindent
where $E_{e^{+}e^{-}D_{s}^{\pm}}=\sqrt{|\vec{p}_{\mathrm{tag}}|^{2}+m^{2}_{D_{s}^{+}}}+E_{e^{+}e^{-}}$, $E_{e^{+}e^{-}}$ is the energy of the $e^{+}e^{-}$ pair in the CM frame, and $\vec{p}_{e^{+}e^{-}D_{s}^{\pm}}=\vec{p}_{\mathrm{tag}}+\vec{p}_{e^{+}e^{-}}$ is the corresponding three-momentum. 
If multiple combinations (tagged $D_{s}^{\pm}$ and multiple $e^{+}e^{-}$ pairs) survive the selections for a specific tag mode, the one with the closest $M^{\mathrm{rec}}_{e^{+}e^{-}D_{s}^{\pm}}$ to the $D_{s}^{+}$ nominal mass~\cite{PDG2023} is retained for further study. To suppress the combinatorial backgrounds, $M_{e^{+}e^{-}D_{s}^{\pm}}^{\mathrm{rec}}$ is kept within $(1.93,~2.03)~\mathrm{GeV}/c^{2}$. 
To extract the signal yields, we introduce two variables referring to the invariant mass of $D_s^{*\pm}$ including a resolution correction, $M_{e^{+}e^{-}D_{s}^{\pm\mathrm{tag}}}$ and $M_{e^{+}e^{-}D_{s}^{\pm\mathrm{rec}}}$, defined as
\begin{equation}
    \label{eq:MDsTagee}
    M_{e^{+}e^{-}D_{s}^{\pm\mathrm{tag}}}=M_{e^{+}e^{-}~\mathrm{tag}D_{s}^{\pm}}-M_{\mathrm{tag}D_{s}^{\pm}}+m_{D_{s}^{+}},
\end{equation}

\begin{equation}
    \label{eq:MDsRecee}
    M_{e^{+}e^{-}D_{s}^{\pm\mathrm{rec}}}=M_{e^{+}e^{-}\mathrm{rec}D_{s}^{\pm}}-M_{\mathrm{rec}D_{s}^{\pm}}+m_{D_{s}^{+}},
\end{equation}

\noindent
where $M_{e^{+}e^{-}\mathrm{tag}D_{s}^{\pm}}$, $M_{e^{+}e^{-}\mathrm{rec}D_{s}^{\pm}}$, $M_{\mathrm{tag}D_{s}^{\pm}}$ and $M_{\mathrm{rec}D_{s}^{\pm}}$ are 
\begin{equation}
    \label{eq:MDstee}
    M_{e^{+}e^{-}\mathrm{tag}D_{s}^{\pm}}=\sqrt{(E_{\mathrm{tag}}+E_{e^{+}e^{-}})^{2}-|\vec{p}_{e^{+}e^{-}~\mathrm{tag}}|^{2}},
\end{equation}
\begin{equation}
    \label{eq:MDst}
    M_{\mathrm{tag}D_{s}^{\pm}}=\sqrt{E_{\mathrm{tag}}^{2}-|\vec{p}_{\mathrm{tag}}|^{2}},
\end{equation}
\begin{equation}
    \label{eq:MDsree}
    M_{e^{+}e^{-}\mathrm{rec}D_{s}^{\pm}}=\sqrt{(\sqrt{s}-E_{\mathrm{tag}})^{2}-|\vec{p}_{\mathrm{tag}}|^{2}},
\end{equation}
\begin{equation}
    \label{eq:MDsr}
    M_{\mathrm{rec}D_{s}^{\pm}}=\sqrt{(\sqrt{s}-E_{\mathrm{tag}}-E_{e^{+}e^{-}})^{2}-|\vec{p}_{e^{+}e^{-}~\mathrm{tag}}|^{2}},
\end{equation}
\begin{equation}
    \label{eq:ptagee}
    \vec{p}_{e^{+}e^{-}~\mathrm{tag}}=\vec{p}_{\mathrm{tag}}+\vec{p}_{e^{+}e^{-}}.
\end{equation}

\begin{figure}[!htp]
    \begin{center}
        \includegraphics[width=0.39\textwidth, height=0.45\textheight, trim=5 0 0 0, clip]{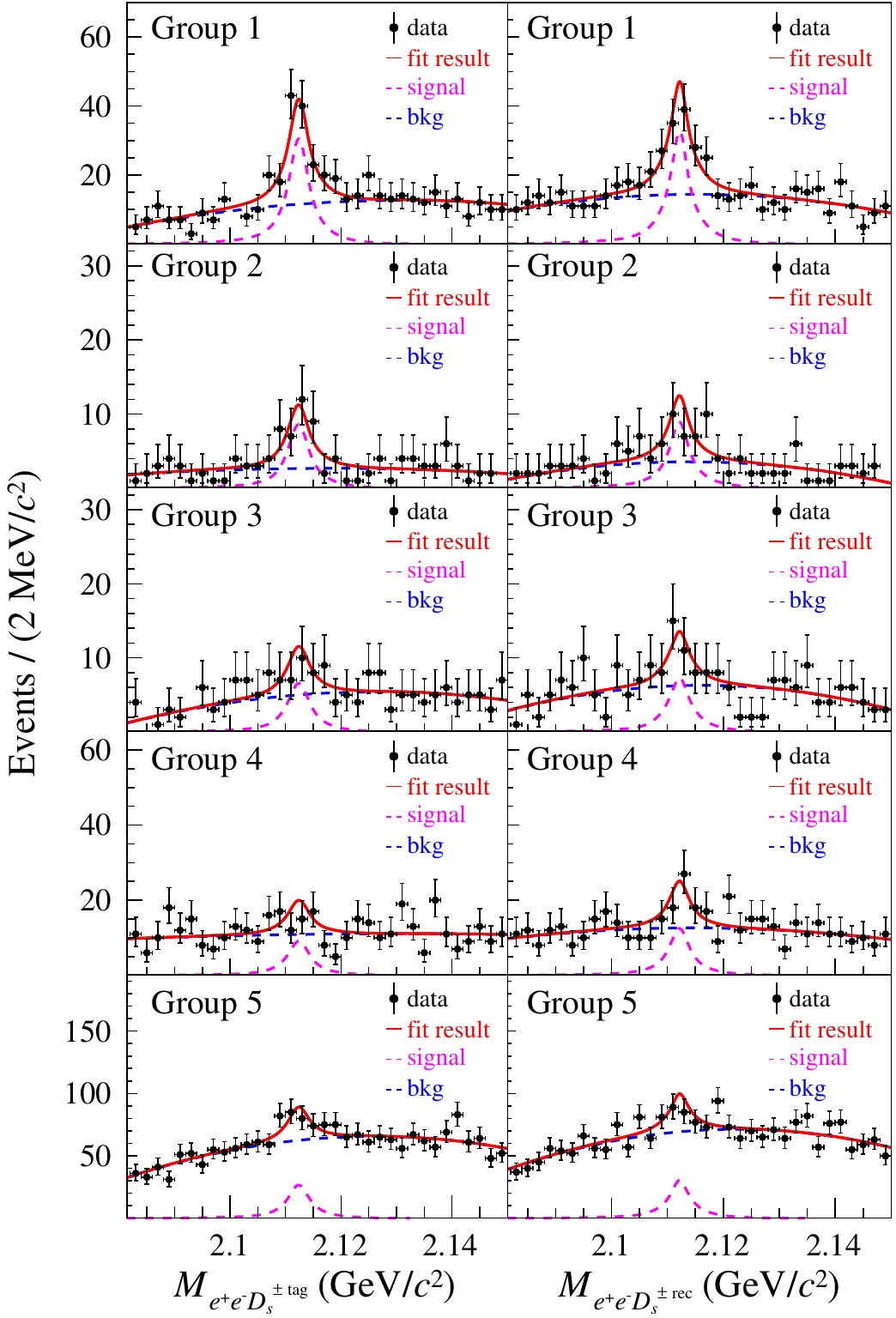}
    \end{center}
    \caption{
      The $M_{e^{+}e^{-}D_{s}^{\pm\mathrm{tag}}}$ and $M_{e^{+}e^{-}D_{s}^{\pm\mathrm{rec}}}$ distributions of the signal candidates by combining the all eight data samples with different CM energies. The black dots with error bars are data. The red solid curves represent the fit results and the blue (pink) dashed curves describe the background (signal) shapes.}
     \label{fig:Sig_AllE}
\end{figure}

Due to limited statistics in some tag modes, the eleven tag modes are grouped into five categories based on similar signal-to-background ($S/B$) ratios where $S$ and $B$ denote the numbers of signal and background events from MC simulated samples, as summarized in Table~\ref{tab:5group}. The distributions of $M_{e^{+}e^{-}D_{s}^{\pm\mathrm{tag}}}$ and $M_{e^{+}e^{-}D_{s}^{\pm\mathrm{rec}}}$ for the five groups of tag modes are shown in Fig.~\ref{fig:Sig_AllE}, where the charge conjugated modes and the data samples collected at different CM energies are summed. The clear $D_{s}^{*\pm}$ signals observed in the $M_{e^{+}e^{-}D_{s}^{\pm\mathrm{tag}}}$ distribution indicate that the tagged $D_s^{\pm}$ is the daughter one, while those observed in the $M_{e^{+}e^{-}D_{s}^{\pm\mathrm{rec}}}$ distribution indicate the tagged $D_s^{\pm}$ as the bachelor one.

\begin{table}[htbp]
	\centering
	\caption{Summary of the five groups of tag modes and the corresponding signal-to-background ratios based on the inclusive MC sample.}	
	\label{tab:5group}
	
	\begin{tabular}[b]{ l |  c }
		\hline
		\hline
        \makebox[0.070\textwidth][l]{\small{Group}} & \makebox[0.38\textwidth][c]{\small{Tag mode ($S/B$)}} \\
		\hline
        Group 1 & $K^{+}K^{-}\pi^{\pm}~(1.15)$ \\
        \hline

        Group 2 & $K_{S}^{0}K^{\pm}~(1.82)$, $\pi^{\pm}\eta'_{\pi^{+}\pi^{-}\eta}~(0.99)$ \\
        \hline

        Group 3 & $K_{S}^{0}K^{\mp}\pi^{\pm}\pi^{\pm}~(0.77)$, $\pi^{\pm}\eta~(0.64)$ \\
        \hline

        Group 4 & $K_{S}^{0}K^{\pm}\pi^{0}~(0.42)$, $K^{+}K^{-}\pi^{\pm}\pi^{0}~(0.48)$ \\
        \hline

        Group 5 & \makecell{$\pi^{+}\pi^{-}\pi^{\pm}~(0.25)$, $\pi^{\pm}\pi^{0}\eta~(0.20)$,\\ $\pi^{\pm}\eta'_{\gamma\rho^{0}}~(0.27)$, $K^{\pm}\pi^{+}\pi^{-}~(0.18)$} \\
        
		\hline	
		\hline
	\end{tabular}
\end{table}

Based on Eqs.~\ref{eq:STY}~and~\ref{eq:DTY}, the branching fraction of $D_{s}^{*+}\to e^{+}e^{-}D_{s}^{+}$ is extracted as
\begin{equation}
    \label{eq:SigBr2}
    \mathcal{B}(g)=\frac{2N(g, f^{\pm, \mathrm{dau/bac}})}{S(f^{\pm})\times\epsilon(g, f^{\pm, \mathrm{dau/bac}})/\epsilon(f^{\pm})},
\end{equation}
\noindent
where the branching fraction $\mathcal{B}(g)$ can be extracted by the different tag modes and charged conjugated modes, individually. Combining the charged conjugated modes and some of the tag modes, the branching fraction can be extracted as  
\begin{equation}
    \label{eq:SigBr}
    \mathcal{B}(g)=\frac{2~\sum_{f} N(g, f^{\mathrm{dau/bac}})}{\sum_{f}(S(f^{+})+ S(f^{-}))\times\epsilon_{\mathrm{eff}}^{\mathrm{dau/bac}}(g,f)},
\end{equation}
\noindent
with
\begin{equation}
    \label{eq:effBr}
    \epsilon_{\mathrm{eff}}^{\mathrm{dau/bac}}(g,f)=\frac{\epsilon(g, f^{+,\mathrm{dau/bac}})+\epsilon(g, f^{-,\mathrm{dau/bac}})}{\epsilon(f^{+})+\epsilon(f^{-})}.
\end{equation}

To extract the branching fraction of $D^{*\pm}_{s}\to e^+e^- D_{s}^{\pm}$ directly, a 
simultaneous unbinned maximum likelihood fit is performed on the distributions of $M_{e^{+}e^{-}D_{s}^{\pm\mathrm{tag}}}$ and $M_{e^{+}e^{-}D_{s}^{\pm\mathrm{rec}}}$ for the five groups of tag modes and the eight data samples with different CM energies (40 sub-data samples in total). 

In the fit, the signal shapes are described by those from MC simulation convolved with a Gaussian function to account for the difference in resolution between data and MC simulation, where the parameters of this Gaussian function are shared in all sub-data samples. 
The backgrounds from continuum hadron production in the $e^{+}e^{-}$ annihilation and those from open charm processes 
are described with a second order Chebychev polynomial function, whose parameters are also shared among the sub-data samples with different CM energies, but are treated separately among the five tag mode groups.
The signal yield in each sub-data sample is constrained by the common branching fraction of $D_{s}^{*\pm}\to e^+e^- D_{s}^{\pm}$, as defined in Eq.~\ref{eq:SigBr}, which is shared among all sub-data samples, while the background yields are treated as free and independent parameters for each sub-data sample.
Additionally, detailed studies based on the signal MC samples indicate a plateau in the $M_{e^{+}e^{-}D_{s}^{\pm\mathrm{tag}}}$ and $M_{e^{+}e^{-}D_{s}^{\pm\mathrm{rec}}}$ distributions due to wrong combinations.
The effects are also considered in the simultaneous fit, where the shapes are derived from the signal MC sample, and their relative normalizations are constrained to the same ratios with respect to the signal as obtained from the signal MC.

In the simultaneous unbinned maximum likelihood fit, the yields of each tag mode ($S(f^\pm)$) as well as their detection efficiencies $\epsilon (f^\pm)$, and the signal detection efficiencies $\epsilon(g, f^{\pm,\mathrm{dau/bac}})$ obtained from the signal MC samples are summarized in Tables~\ref{tab:STDspm} and~\ref{tab:Sigeff} for the data sample at $\sqrt{s}=4.178$~GeV, respectively. For the other data samples with different CM energies, they are presented in the Supplemental Material~\cite{Supp}.
The fit curves for the five groups of tag modes summing over the data samples at different CM energies are shown in Fig.~\ref{fig:Sig_AllE}, where good fit quality is obtained.
The fit results in the branching fraction ${\mathcal{B}(D^{*+}_{s}\to e^{+}e^{-}D^{+}_{s})=(7.28\pm0.61)\times10^{-3}}$; the total signal yield is calculated as $573\pm48$ with Eq.~\ref{eq:SigBr}, where the uncertainties are statistical only.

\begin{table*}[htbp]
	\centering
	\caption{Summary of the signal efficiencies $\epsilon(g, f^{\pm,\mathrm{dau/bac}})$ (\%) for individual ST modes at $\sqrt{s}=4.178~\mathrm{GeV}$. The uncertainties are statistical only.}	
	\label{tab:Sigeff}
    \begin{tabular}[b]{ l  p{1.2cm}<{\raggedleft} @{ $\pm$ } p{1.2cm}<{\raggedright} p{1.2cm}<{\raggedleft} @{ $\pm$ } p{1.2cm}<{\raggedright} p{1.2cm}<{\raggedleft} @{ $\pm$ } p{1.2cm}<{\raggedright} p{1.2cm}<{\raggedleft} @{ $\pm$ } p{1.2cm}<{\raggedright} }
        \hline
        \hline
        \makebox[0.065\textwidth][l]{\small{Mode}} & \multicolumn{2}{c}{$\epsilon(g,f^{+,\mathrm{dau}})$} & \multicolumn{2}{c}{$\epsilon(g,f^{+,\mathrm{bac}})$} & \multicolumn{2}{c}{$\epsilon(g,f^{-,\mathrm{dau}})$} & \multicolumn{2}{c}{$\epsilon(g,f^{-,\mathrm{bac}})$} \\

        \hline
        
        $K_{S}^{0}K^{\pm}$                    & 4.57 & 0.08 & 4.77 & 0.08 & 4.58 & 0.10 & 4.90 & 0.08 \\

        $K^{+}K^{-}\pi^{\pm}$                 & 3.88 & 0.03 & 4.17 & 0.03 & 3.90 & 0.03 & 4.15 & 0.03 \\

        $K_{S}^{0}K^{\pm}\pi^{0}$             & 1.46 & 0.04 & 1.86 & 0.05 & 1.47 & 0.04 & 1.77 & 0.05 \\

        $K^{+}K^{-}\pi^{\pm}\pi^{0}$          & 0.93 & 0.02 & 1.26 & 0.02 & 0.96 & 0.02 & 1.25 & 0.02 \\

        $K_{S}^{0}K^{\mp}\pi^{\pm}\pi^{\pm}$  & 1.67 & 0.04 & 1.95 & 0.05 & 1.71 & 0.05 & 1.91 & 0.05 \\

        $\pi^{+}\pi^{-}\pi^{\pm}$             & 5.41 & 0.08 & 5.88 & 0.09 & 5.29 & 0.08 & 5.94 & 0.09 \\

        $\pi^{\pm}\eta$                       & 4.95 & 0.10 & 5.22 & 0.11 & 4.88 & 0.10 & 5.19 & 0.11 \\

        $\pi^{\pm}\pi^{0}\eta$                & 2.17 & 0.03 & 2.59 & 0.03 & 2.17 & 0.03 & 2.48 & 0.03 \\

        $\pi^{\pm}\eta'_{\pi^{+}\pi^{-}\eta}$ & 2.19 & 0.05 & 2.35 & 0.06 & 2.16 & 0.05 & 2.41 & 0.06 \\

        $\pi^{\pm}\eta'_{\gamma\rho^{0}}$     & 3.12 & 0.08 & 3.23 & 0.08 & 3.14 & 0.08 & 3.20 & 0.08 \\

        $K^{\pm}\pi^{+}\pi^{-}$               & 4.56 & 0.10 & 5.07 & 0.11 & 4.67 & 0.11 & 4.94 & 0.11 \\

		\hline	
		\hline
	\end{tabular}
\end{table*}

\section{SYSTEMATIC UNCERTAINTIES STUDIES}
\label{sec:syst_uncer}

All sources of systematic uncertainty in the branching fraction measurement are discussed below, and summarized in Table~\ref{tab:Syst_br}.

\begin{table}[htbp]
	\centering
	\caption{Summary of the systematic uncertainties for the branching fraction measurement.}	
	\label{tab:Syst_br}
    \begin{tabular}[b]{ l c }
        \hline
        \hline
        \makebox[0.140\textwidth][l]{\small{Source}} & \makebox[0.240\textwidth][c]{\small{Systematic uncertainty~(\%)}} \\
        \hline
        \small{ST} & $0.4$ \\
        \small{Tracking of $e^{+}e^{-}$ pair} & $1.7$ \\
        \small{PID of $e^{+}e^{-}$ pair} & $1.4$ \\
        \small{$\gamma$ conversion rejection} & $2.8$ \\
        \small{Signal shape} & $1.9$ \\
        \small{Background shape} & $1.2$ \\
        \small{Fitting region} & $1.1$ \\
        \hline
        \small{Total} & $4.3$ \\
		\hline	
		\hline
	\end{tabular}
\end{table}

From Eqs.~\ref{eq:SigBr} and~\ref{eq:effBr}, it can be seen that the systematic uncertainties in the branching fraction measurement arise from contributions associated with the tagged and signal yields, as well as from the ratio of detection efficiencies between the signal and tagged samples. 

The uncertainty associated with the single tag yields is 0.4\%, which includes the statistical uncertainty and a systematic component estimated by performing the fits without convolving the Gaussian function on the signal shape, and changing the background shape from a second order Chebychev polynomial to a third order; the resultant differences with respect to the nominal values are taken as the uncertainties. Similarly, the uncertainty associated with the signal yields is estimated by changing the left and right boundaries of the fit range by $2~\mathrm{MeV}/c^{2}$, as well as the shapes of the signal and the background; this contribution amounts to 2.5\%, assuming these sources of uncertainties uncorrelated and taking the quadratic sum as the total uncertainty.

The systematic uncertainties from detection efficiencies are canceled for the single tag efficiencies, according to Eqs.~\ref{eq:SigBr} and~\ref{eq:effBr}. The uncertainties associated with $e^{+}e^{-}$ pair include the contributions from the $e^{+}e^{-}$ tracking, PID and photon conversion rejection. The uncertainty from tracking (PID) is determined by studying a control sample $J/\psi\to\pi^{+}\pi^{-}\pi^{0}$, $\pi^{0}\to e^{+}e^{-}\gamma$, 
and the signal detection efficiencies are reweighted according to the two-dimensional kinematic distribution of transverse momentum (momentum) and $\cos\theta$. The overall correction factor is estimated to be $(123.3\pm2.1)\%$ (($90.6\pm1.2)\%$). The residual uncertainties are assigned as the $e^{+}e^{-}$ pair tracking (PID) systematic uncertainties after correcting the signal efficiencies. The uncertainty from the photon conversion rejection is determined by the same control sample; the signal efficiency is corrected by the ratio $(105.4\pm2.9)\%$, which is the vetoing photon conversion efficiency between data and MC simulation, and the residual uncertainty is assigned as the photon conversion rejection systematic uncertainty. The uncertainties associated with the $e^{+}e^{-}$ tracking, PID and photon conversion rejection are assigned to be 1.7\%, 1.4\% and 2.8\%, respectively. All other uncertainties are negligible. Assuming that all the sources of uncertainties are uncorrelated, the quadratic sum of the individual values, 4.3\%, is taken as the total uncertainty.

\section{SUMMARY}

In summary, the branching fraction of the EM Dalitz decay $D_{s}^{*+}\to e^{+}e^{-}D_{s}^{+}$ is measured by using an electron-positron collision data sample corresponding to a total integrated luminosity of $7.33~\mathrm{fb}^{-1}$, collected at CM energies between 4.128~GeV and 4.226~GeV with the BESIII detector. The measured branching fraction ${\mathcal{B}(D^{*+}_{s}\to e^{+}e^{-}D^{+}_{s})=(7.28\pm0.61_{\mathrm{stat}}\pm0.31_{\mathrm{syst}})\times10^{-3}}$ is consistent with the previous CLEO-c result of ${(6.7^{+1.4}_{-1.2}\pm0.9)\times10^{-3}}$, and the precision is improved by a factor of 2.5. Taking the branching fraction of $D_{s}^{*+}\to\gamma D_{s}^{+}$ from the PDG, $\mathcal{B}(D_{s}^{*+}\to\gamma D_{s}^{+})=0.936\pm0.004$~\cite{PDG2023}, the ratio of the branching fractions between $D_{s}^{*+}\to e^{+}e^{-}D_{s}^{+}$ and $D_{s}^{*+}\to\gamma D_{s}^{+}$ is calculated to be $(7.78\pm0.73)\times10^{-3}$, which is consistent with the theoretical prediction of $6.5\times10^{-3}$~\cite{EMD_Ds} within $1.8\sigma$.

\section{ACKNOWLEDGEMENT}

\input{./acknowledgement_2025-08-14.tex}

\bibliographystyle{./apsrev4-1}
\bibliography{bibitem}

\nolinenumbers

\end{document}

%% file: authorlist_2025-08-14.tex
M.~Ablikim$^{1}$\BESIIIorcid{0000-0002-3935-619X},
M.~N.~Achasov$^{4,b}$\BESIIIorcid{0000-0002-9400-8622},
P.~Adlarson$^{81}$\BESIIIorcid{0000-0001-6280-3851},
X.~C.~Ai$^{86}$\BESIIIorcid{0000-0003-3856-2415},
R.~Aliberti$^{39}$\BESIIIorcid{0000-0003-3500-4012},
A.~Amoroso$^{80A,80C}$\BESIIIorcid{0000-0002-3095-8610},
Q.~An$^{77,64,\dagger}$,
Y.~Bai$^{62}$\BESIIIorcid{0000-0001-6593-5665},
O.~Bakina$^{40}$\BESIIIorcid{0009-0005-0719-7461},
Y.~Ban$^{50,g}$\BESIIIorcid{0000-0002-1912-0374}
% ...
\\
\vspace{0.2cm}
(BESIII Collaboration)\\
\vspace{0.2cm} {\it
$^{1}$ Institute of High Energy Physics, Beijing 100049, People's Republic of China\\
% ...
$^{4}$ Budker Institute of Nuclear Physics SB RAS (BINP), Novosibirsk 630090, Russia\\
% ...
$^{39}$ Johannes Gutenberg University of Mainz, Johann-Joachim-Becher-Weg 45, D-55099 Mainz, Germany\\
$^{40}$ Joint Institute for Nuclear Research, 141980 Dubna, Moscow region, Russia\\
% ...
$^{50}$ Peking University, Beijing 100871, People's Republic of China\\
% ...
$^{62}$ Southeast University, Nanjing 211100, People's Republic of China\\
% ...
$^{64}$ State Key Laboratory of Particle Detection and Electronics, Beijing 100049, Hefei 230026, People's Republic of China\\
% ...
$^{77}$ University of Science and Technology of China, Hefei 230026, People's Republic of China\\
% ...
$^{80}$ University of Turin and INFN, (A)University of Turin, I-10125, Turin, Italy; (B)University of Eastern Piedmont, I-15121, Alessandria, Italy; (C)INFN, I-10125, Turin, Italy\\
$^{81}$ Uppsala University, Box 516, SE-75120 Uppsala, Sweden\\
% ...
$^{86}$ Zhengzhou University, Zhengzhou 450001, People's Republic of China\\

\vspace{0.2cm}
$^{\dagger}$ Deceased\\
% ...
$^{b}$ Also at the Novosibirsk State University, Novosibirsk, 630090, Russia\\
% ...
$^{g}$ Also at State Key Laboratory of Nuclear Physics and Technology, Peking University, Beijing 100871, People's Republic of China\\
% ...

}

%% file: acknowledgement_2025-08-14.tex

The BESIII Collaboration thanks the staff of BEPCII (https://cstr.cn/31109.02.BEPC), the IHEP computing center and the supercomputing center of the University of Science and Technology of China (USTC) for their strong support. The authors are grateful to Doctor Xiao-Dong Shi and Professor Dao-Neng Gao for enlightening discussions. This work is supported in part by National Key R\&D Program of China under Contracts Nos. 2023YFA1606000, 2023YFA1606704, 2023YFA1609400; National Natural Science Foundation of China (NSFC) under Contracts Nos. 11635010, 11935015, 11935016, 11935018, 12025502, 12035009, 12035013, 12061131003, 12105276, 12122509, 12192260, 12192261, 12192262, 12192263, 12192264, 12192265, 12221005, 12225509, 12235017, 12361141819, 12475091; Guangzhou Navigation Project No. 2024A04J6334; the Chinese Academy of Sciences (CAS) Large-Scale Scientific Facility Program; the Strategic Priority Research Program of Chinese Academy of Sciences under Contract No. XDA0480600; CAS under Contract No. YSBR-101; Joint Large-Scale Scientific Facility Funds of the NSFC and CAS under Contracts Nos. U2032111; 100 Talents Program of CAS; The Institute of Nuclear and Particle Physics (INPAC) and Shanghai Key Laboratory for Particle Physics and Cosmology; ERC under Contract No. 758462; German Research Foundation DFG under Contract No. FOR5327; Istituto Nazionale di Fisica Nucleare, Italy; Knut and Alice Wallenberg Foundation under Contracts Nos. 2021.0174, 2021.0299; Ministry of Development of Turkey under Contract No. DPT2006K-120470; National Research Foundation of Korea under Contract No. NRF-2022R1A2C1092335; National Science and Technology fund of Mongolia; Polish National Science Centre under Contract No. 2024/53/B/ST2/00975; STFC (United Kingdom); Swedish Research Council under Contract No. 2019.04595; U. S. Department of Energy under Contract No. DE-FG02-05ER41374.
